\documentclass[12pt,preprint]{aastex}

\newcommand\kmix{\kappa_{\rm mix}}
\newcommand\kcond{\kappa_{\rm cond}}
\newcommand\amix{\alpha_{\rm mix}}

\newcommand\kms{\,{\rm km\,s^{-1}}}
\newcommand\vturb{v_{\rm turb}}
\newcommand\kpcMyr{\,{\rm kpc^2\,Myr^{-1}}}
\newcommand\dpot{\Phi}
\newcommand\cs{c_s}
\newcommand\cm{{\rm cm}}
\newcommand\simgt{\lower.5ex\hbox{$\; \buildrel > \over \sim \;$}}
\newcommand\simlt{\lower.5ex\hbox{$\; \buildrel < \over \sim \;$}}

\slugcomment{Accepted for Publication in ApJL}

\shortauthors{Kim \& Narayan}
\shorttitle{Turbulent Mixing in Galaxy Clusters}

\begin{document}

\title{Turbulent Mixing in Clusters of Galaxies}

\author{Woong-Tae Kim and Ramesh Narayan}
\affil{Harvard-Smithsonian Center for Astrophysics, \\
60 Garden Street, Cambridge, MA 02138}

\email{wkim@cfa.harvard.edu, rnarayan@cfa.harvard.edu}

\begin{abstract}
We present a spherically-symmetric, steady-state model of galaxy
clusters in which radiative cooling from the hot gas is balanced by
heat transport through turbulent mixing.  We assume that the gas is in
hydrostatic equilibrium, and describe the turbulent heat diffusion by
means of a mixing length prescription with a dimensionless parameter
$\amix$.  Models with $\amix\sim0.01-0.03$ yield reasonably good fits
to the observed density and temperature profiles of cooling core
clusters.  Making the strong simplification that $\amix$ is
time-independent and that it is roughly the same in all clusters, the
model reproduces remarkably well the observed scalings of X-ray
luminosity, gas mass fraction and entropy with temperature.  The break
in the scaling relations at $kT\sim1-2$ keV is explained by the break
in the cooling function at around this temperature, and the entropy
floor observed in galaxy groups is reproduced naturally.
\end{abstract}

\keywords{galaxies: clusters --- cooling flows --- X-rays: galaxies ---
turbulence --- hydrodynamics --- instability }

\section{Introduction}

Recent X-ray observations \citep[and references therein]{pet01,pet03}
show that there must be some heat source balancing the radiative
cooling of hot gas in galaxy clusters.  Current ideas include (1) heat
transport from the outer regions of the cluster to the center by
thermal conduction (e.g.,
\citealt{ber86,bre88,ros89,nar01,dos01,voi02,zak03}, hereafter ZN03),
(2) energy injection by jets or radiation from a central active
galactic nucleus (AGN; \citealt{chu00,chu02,cio01,bru02,kai03}), or
(3) a combination of both (e.g., ZN03,
\citealt{rus02,bri02,bri03,bru03}).

Models based on electron conduction are able to reproduce the observed
profiles of gas density and temperature in several clusters (ZN03).
The models are also much less unstable than models without conduction
\citep[][hereafter KN03]{kim03}.  These features make the conduction
model rather attractive.  In addition to conduction, another diffusive
process that may be potentially important in galaxy clusters is
turbulent mixing.  Clusters are highly dynamic entities that are
constantly being stirred by the infall of groups and subclusters, the
motions of galaxies, and outflows from AGN.  The dynamical motions
supply a large amount of turbulent kinetic energy, perhaps comparable
to the thermal energy of the cluster (e.g.,
\citealt{dei96,roe99,ric01}).  The turbulence also causes diffusive
mixing which tends to erase gradients in the specific entropy and
chemical composition \citep{sar88} and transports heat efficiently
\citep{cho03}.  The latter feature of turbulence is the focus of this
Letter.  We describe a model of galaxy clusters in which turbulent
heat transport balances radiative cooling of the hot gas, and we show
that this model is consistent with a wide range of observations.

\section{Model}

In the presence of turbulence, gas elements move around randomly and
transport specific entropy from one point to another.  When the
elements mix, there is a net heat flux from regions of higher entropy
to regions of lower entropy.  The heat flux due to turbulence is thus
proportional to the entropy gradient and may be written as
\begin{equation}\label{tflux}
\mathbf{F} = -\kmix \rho T\nabla s, \quad \kmix = \amix\cs H_P, \quad
H_P \approx (r^2+r_c^2)^{1/2},
\end{equation}
where $\rho$ is the gas density, $T$ is the temperature, and $s$ is
the entropy per unit mass.  The diffusion constant $\kmix$ is expected
to be of order $\vturb l_{\rm coh}/3$, where $\vturb$ is the rms
turbulent velocity and $l_{\rm coh}$ is the coherence length of the
velocity field \citep[e.g.,][]{dei96,cho03}.  We expect $v_{\rm turb}$
to be a fraction of the sound speed $c_s$ (the only relevant velocity
in the problem) and $l_{\rm coh}$ to be a fraction of the local
pressure scale height $H_P$ (the obvious scale, as in the mixing
length theory of convection).  Over most of the cluster, the scale
$H_P$ is comparable to the radius, but it is of order the core radius
$r_c$ inside the core (see ZN03 for the definition of $r_c$).  These
arguments lead to the scalings for $\kmix$ and $H_P$ shown in equation
(\ref{tflux}).  The specific entropy for a classical perfect gas is
given by $s=c_v\ln (P\rho^{-\gamma})$, where $P=\rho c_s^2$ is the
pressure, $c_v$ is the specific heat at constant volume, and
$\gamma=5/3$ is the adiabatic index of the gas.

We consider a spherically symmetric cluster in hydrostatic
equilibrium, $(1/\rho)dP/dr = - d\dpot/dr$, where $\dpot$ is the
gravitational potential.  We assume steady state, so that the
radiative cooling of each gas element is balanced by the divergence of
$F$:
\begin{equation}\label{engy}
\nabla\cdot\mathbf{F} = -j,
\end{equation}
where $j$ is the radiative energy loss rate per unit volume. For
$T\simgt2$ keV, $j$ is dominated by thermal bremsstrahlung from free
electrons and ions, while for lower temperatures it is mostly due to
atomic transitions.  We should note that the assumption of steady
state is a strong simplification.  Since turbulent motions in a
cluster are likely to be episodic, e.g., as a result of AGN activity
\citep{kai03} or sub-cluster infall and merger \citep{dei96,ric01},
equations (\ref{tflux}) and (\ref{engy}) should be viewed as
time-averaged relations.

\section{Results}

Using the expressions for $\Phi$ and $j$ given in ZN03, we have solved
the basic equations described above to calculate the radial profiles
of the electron number density $n_e(r)$ and temperature $T(r)$.  For
each cluster, we selected different values for the the inner gas
density $n_e(0)$ and inner temperature $T(0)$ as well as the mixing
parameter $\amix$, and integrated the steady state equations
numerically.  We compared the calculated profiles of density and
temperature with the data and computed the $\chi^2$ of the fit.  In
those cases where the published data had no error estimates, we
arbitrarily assumed that the errors are 15\% of the measured values.
Also, in some clusters we pruned the density data so as to have equal
numbers of data points in the density and temperature profiles.  By
minimizing $\chi^2$, we determined the optimum values of $n_e(0)$,
$T(0)$ and $\alpha_{\rm mix}$ for each cluster.

\begin{figure}
\epsscale{1.}
\plottwo{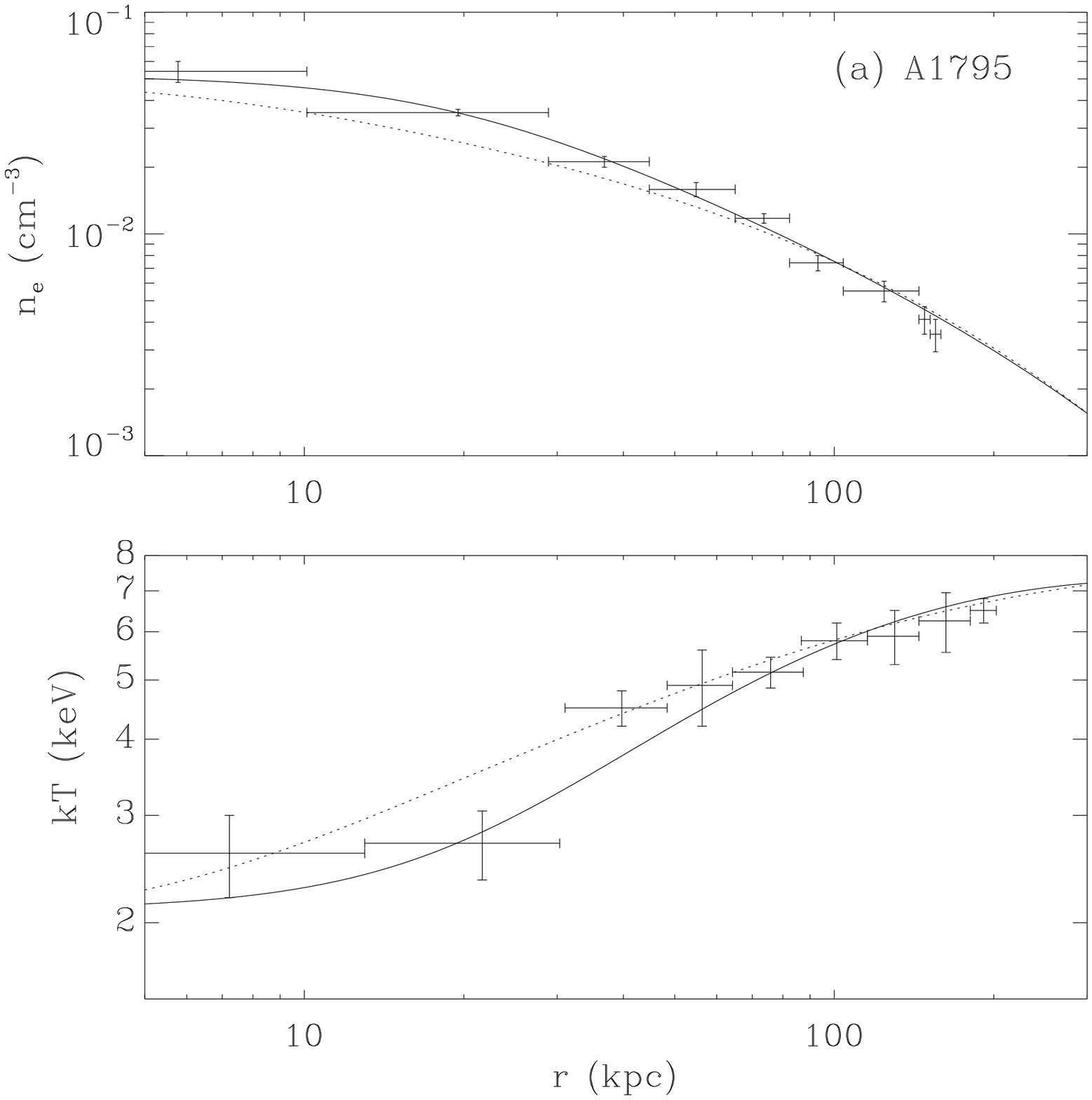}{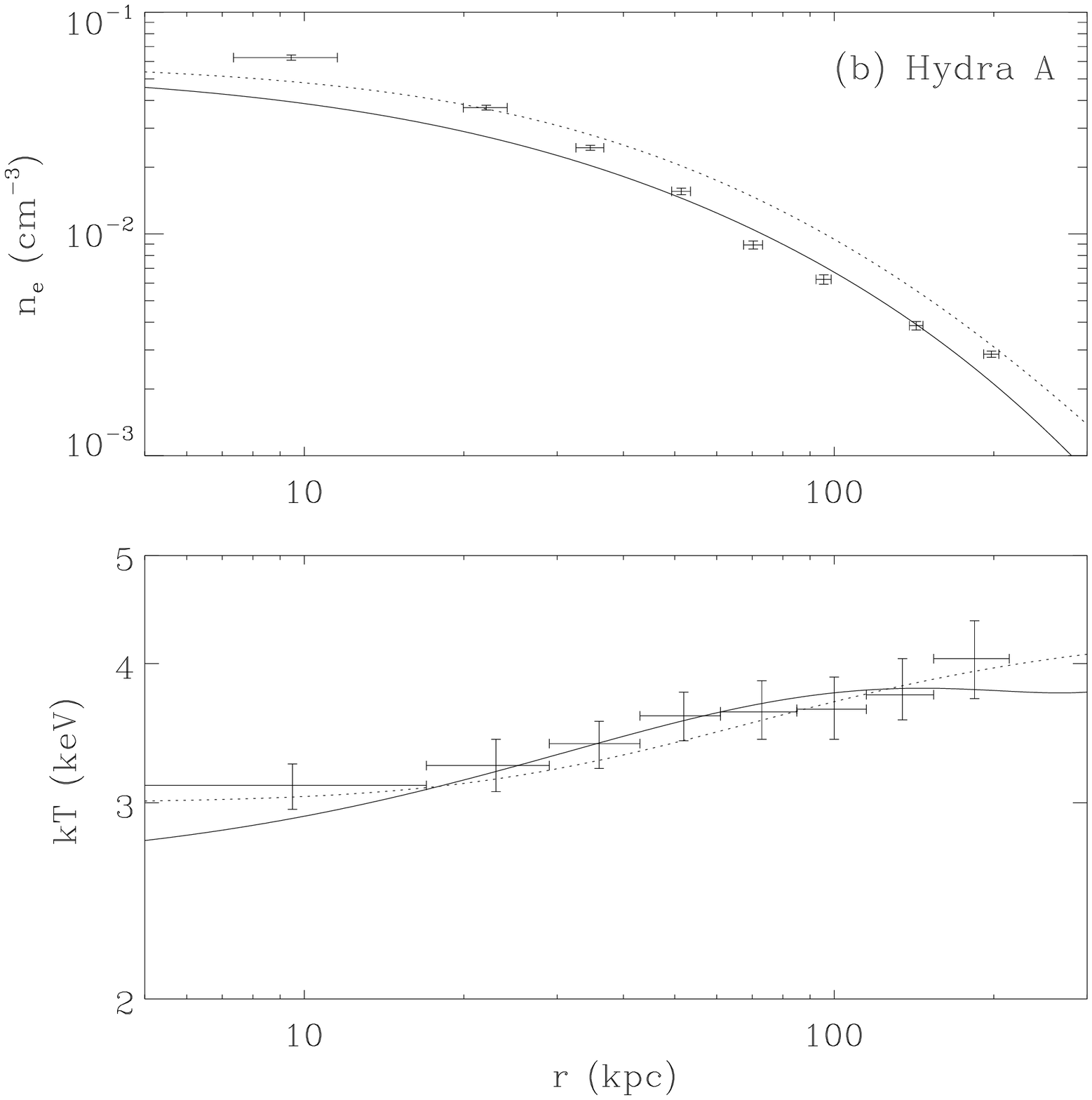}
\caption{Observed and calculated density and temperature profiles for
(a) A1795 and (b) Hydra A.  Results for the turbulent mixing model are
shown by solid lines and those for the conduction model are shown by
dotted lines.  Crosses correspond to {\it Chandra} data
(\citealt{ett02} for A1795 and \citealt{dav01} for Hydra A).
$H_0=70\,{\rm km\,s^{-1}\,Mpc^{-1}}$, $\Omega_M=0.3$, and
$\Omega_\Lambda=0.7$ are adopted.  The parameters for the mixing model
are given in Table \ref{tbl1}.  The parameters for the conduction model 
are $f \equiv \kappa_{\rm cond}/\kappa_{\rm Spitzer} = 0.2$,
$n_e(0)=0.049\,\cm^{-3}$, $T(0)=2.0\,$keV for A1795, and $f=3.5$,
$n_e(0)=0.06\,\cm^{-3}$, $T(0)=3.0\,$keV for Hydra A.  Because $f>1$
for Hydra A, the conduction model is not viable for this cluster
(ZN03).
\label{fig1}}
\end{figure}

Table \ref{tbl1} shows the results of fitting the turbulent mixing
model to the ten clusters analyzed by ZN03.  ZN03 found that five of
these clusters could be explained with the thermal conduction model,
while five required unphysically high levels of conduction and were
inconsistent.  With the turbulent mixing model, however, we find that
we obtain fairly good fits for all ten clusters, with reasonably small
(and physically justifiable) values of $\alpha_{\rm mix}$.  The five
clusters (A1795, A1835, A2199, A2390, and RX J1347.5-1145) that ZN03
found to be consistent with conduction have a median $\amix$ of 0.013,
while the other five clusters (A2052, A2597, Hydra A, Ser 159-03, and
3C 295) have a median $\amix$ of 0.026.  The larger $\amix$ in the
latter group may reflect the fact that these clusters have powerful
AGN that cause extra turbulence.  Figure \ref{fig1} shows the results
of the model-fitting for two clusters, A1795 (one of the
better-fitting clusters) and Hydra A (the worst example).  We see that
the mixing model is generally consistent with the observations, though
the $\chi^2$ is not always small.

Apart from providing a qualitative explanation for the observed
density and temperature profiles of cluster gas, another major
attraction of the thermal conduction model of ZN03 is the fact that
conduction helps to control the thermal instability of the gas (KN03).
We have now repeated the global stability analysis of KN03 for the
turbulent mixing model.  For A1795, with $\amix=0.011$, we find that
mixing suppresses the instability in all radial modes except the
fundamental (nodeless) mode.  The growth time of the lone unstable
mode is very much longer than the Hubble time.  In the case of Hydra
A, with $\amix=0.021$ (and thus a larger $\kmix$), we find that all
modes, including the nodeless fundamental mode, are stable.  Thus, the
equilibrium models described here are for all practical purposes
stable.

\section{Scaling Relations}

Groups and clusters of galaxies form when primordial density
perturbations in the universe grow, gravitationally collapse, and
merge together according to the standard hierarchical clustering
scenario.  The statistical properties of these collapsed systems
contain many clues to the process of cosmic structure formation.
Numerous X-ray studies have been published on the power-law scalings
of the size, temperature, X-ray luminosity, mass, entropy, and gas
mass fraction of galaxy clusters. The best fit values of the power-law
indices depend on the particular sample of clusters used and on the
specific methods employed to estimate the mass and temperature.
Nevertheless, there is a broad consensus on the observed scalings as
functions of temperature $T$, as summarized in the first three columns
of Table \ref{tbl2}.  A clear break in cluster properties is seen at a
characteristic temperature $\sim1-2$ keV.

Small clusters and galaxy groups are found to have a relatively constant
entropy.  The prevailing explanations for this surprising ``entropy
floor'' include (1) pre-heating of intracluster gas
\citep{kai91,evr91} via galactic winds \citep{loe00}, AGN
\citep{val99,wu02,nat02}, and accretion shocks
\citep{dav01,toz01,dos02}, (2) removal of cold, low-entropy gas via
galaxy formation in clusters \citep{bry00,mua01,wu02,dav02}, and (3)
both radiative cooling and supernova feedback \citep{voit01,voit02}.
None of these models includes the effects of thermal conduction or
turbulent mixing.

When thermal conduction by electrons is the dominant heating
mechanism, the local heat flux is given by $\mathbf{F} = -\kcond
\nabla T$, with $\kcond \propto T^{5/2}$ \citep{spi62}.  For $kT>2$
keV, we may assume $j\propto n_e^2T^{1/2}$, corresponding to thermal
bremsstrahlung.  Then, using equation (\ref{engy}) and substituting
$\nabla \sim 1/r_s \sim 1/T^{1/2}$, we find that the gas density in
equilibrium scales as $n_e \sim T$, which gives a gas mass $M_g \sim
n_e r_s^3 \sim T^{5/2}$, gas fraction $f_g \equiv M_g/M \sim T^{0.7}$,
X-ray luminosity $L_X \sim j r_s^3 \sim T^4$, and entropy $S\equiv\exp
(s/c_v) \sim T/n^{2/3} \sim T^{1/3}$.  For the mixing model, on the
other hand, equation (\ref{tflux}) with $c_s \sim T^{1/2}$ and $l\sim
r\sim T^{1/2}$, combined with equation (2), yields $n_e \sim T^{1/2}$,
$M_g \sim T^2$, $f_g \sim T^{0.2}$, $L_X \sim T^3$, and $S \sim
T^{2/3}$.  These scaling relations are shown in the last two columns
of Table \ref{tbl2}.  Remarkably, the turbulent mixing model yields
scaling relations for $kT > 2$ keV that are in very good agreement
with observations.  The scalings obtained with the conduction model
agree less well.

At lower temperatures, the resonance radiation of highly ionized
metals like O, Si, and Fe are responsible for most of the cooling.
From the equilibrium cooling functions provided by \citet{sut93}, we
approximately obtain $j\propto n_e^2T^{-0.7\sim-1}$ for $kT\sim0.05-1$
keV (for a plasma with solar to half-solar metal abundance).
Following the same steps as above, this gives a different set of
scaling relations, as shown in Table \ref{tbl2}.  Both the conduction
and the mixing model predict a dramatic change in the scaling below
$\sim2$ keV, in agreement with the observations.  The entropy floor
that is observed in small clusters and groups is also reproduced
naturally in the models.

\section{Discussion}

Diffusive processes are quite attractive as a heat source in galaxy
clusters since they transport energy to the cluster centers from the
outside, and also help to control the thermal instability.  In this
Letter, we have presented equilibrium models of galaxy clusters in
which the hot gas maintains energy balance between radiative cooling
and heating by turbulent mixing.  To quantify the amount of heat
transported by mixing, we adopt a mixing length prescription in which
the heat flux is proportional to the local gradient of specific
entropy, with a diffusion coefficient parameterized by a dimensionless
constant $\amix$ (eq.\ [\ref{tflux}]).  The mixing model fits the
observed density and temperature profiles of hot gas in clusters
fairly well (\S3, Fig. 1).  The resulting gas configurations are also
very stable.  In both respects, the mixing model performs better than
the conduction model.  This is presumably because turbulent mixing
transports energy more efficiently in the low-temperature high-density
cores of clusters \citep{cho03}.

What is the origin of the turbulence invoked in the mixing model?  It
could be from the infall of subclusters, the motion of the dark matter
potential, the orbital motions of galaxies, or energy input from an
AGN.  \citet{dei96} studied the turbulence generated by motions of
individual galaxies and found that the diffusion constant is
$\kmix\sim(1-10)\kpcMyr$ in the Coma and Perseus clusters.
\citet{ric01} found that the turbulent velocities driven by cluster
mergers are generally subsonic, with $\vturb\sim(0.1-0.3)
c_s\sim100-300\kms$.  It is difficult to measure the coherence length
of the velocity field, but Faraday depolarization measurements suggest
that $l_B\sim5-20$ kpc for magnetic field tangling near the centers of
typical clusters (e.g., \citealt{car02} and references therein).  The
corresponding diffusion coefficient is $\kmix\sim (0.5-6)\kpcMyr$.  In
comparison, the values of $\amix\sim0.01-0.03$ required in the mixing
models presented here correspond to $\kmix\sim(1-6)\kpcMyr$ for
$r\sim50-300$ kpc.  These values are in good agreement with the other
independent estimates.

As shown in \S3, some clusters that contain relatively strong AGN have
somewhat larger values of $\amix$ compared to clusters that do not
have obvious AGN activity.  This is natural if jets from the AGN cause
some of the turbulence in the centers of these clusters.  The
influence of AGN jets has been studied by \citet{chu02}, \citet{bru02}
and \citet{kai03} who showed that powerful jets mix gas with different
entropy and thereby greatly reduce the mass deposition rate in cluster
cores.  \citet{bru02}, in particular, showed that the entropy increase
at the cluster center is quite insensitive to the amount of energy
injection by AGN, which is consistent with our result that the values
of $\amix$ in AGN-dominated clusters are within a factor of 2 of those
in clusters without strong AGN.

The model considered in this paper is highly idealized.  In reality,
we expect both turbulent mixing and electron conduction to influence
heat transport in clusters.  Mixing may be inhomogeneous, e.g., a
central AGN may induce more efficient mixing near the cluster center
compared to the outer regions.  Mixing is also likely to be variable
in time.  These effects need to be investigated in more detail.
Another problem is that we have not considered how a cluster got to
the state it is observed in.  \citet{bre88} showed that cluster models
with thermal conduction have two very distinct states, one that is
unstable and develops a cooling flow and one that is stable but nearly
isothermal.  Equilibrium models with cool cores occur only near the
boundary between the two states.  Mixing will probably give similar
results.  It is presently unclear in either model how a cluster that
begins from generic initial conditions can end up in such an
apparently non-generic final state.

Despite the above uncertainties, if we assume that all galaxy groups
and clusters are in thermal steady state (at least in a time-averaged
sense), with a balance between cooling and heating via turbulent heat
transport, and if we further assume that $\amix$ is roughly constant
across the entire population, then the model makes clear predictions
for how various cluster properties should scale as a function of
temperature (\S4, Table 2).  Remarkably, the predicted scalings
obtained with the mixing model are in excellent agreement with
observations.  The model also explains the observed discontinuity in
cluster properties at $kT\sim1-2$ keV as a consequence of the change
in the cooling function at this temperature, and provides a natural
explanation for the entropy floor seen in galaxy groups. These
results, though not inconsistent with some of the previously proposed
explanations (see references in \S4), suggest that those models should
be generalized to include the effect of turbulent heat transport.

After this paper was submitted to the journal, Voigt \& Fabian posted
a paper (astro-ph/0308352) in which they have suggested independently
that turbulent heat transport may be important in galaxy clusters.

\acknowledgements The authors thank Nadia Zakamska and Larry David for
providing the cluster data shown in Figure 1, and the referee for
helpful comments.  This work was supported in part by NASA grant
NAG5-10780 and NSF grant AST 0307433.

\begin{deluxetable}{lcccr}
\tablecaption{Best-fit Parameters of the Turbulent Mixing Model 
\label{tbl1}}
\tablewidth{0pt}
\tablehead{
\colhead{Name}                                                     & 
\colhead{\begin{tabular}{c} $T(0)$   \\ (keV)       \end{tabular}} & 
\colhead{\begin{tabular}{c} $n_e$(0) \\ (cm$^{-3}$) \end{tabular}} & 
\colhead{$\amix$}                                                  &
\colhead{$\chi^2$/dof}
}\startdata
Abell 1795      &2.1    &0.052          &0.011    & 8.6/15 \\
Abell 1835      &3.6    &0.070          &0.014    & 6.2/15 \\
Abell 2199      &1.3    &0.10           &0.013    & 6.8/13 \\
Abell 2390      &2.9    &0.070          &0.012    & 8.5/13 \\
RXJ 1347.5-1145 &4.0    &0.34           &0.025    & 7.5/13 \\
Abell 2052      &1.0    &0.043          &0.014    & 62/29 \\
Abell 2597      &1.4    &0.070          &0.026    & 6.6/19  \\
Hydra A         &2.6    &0.057          &0.021    & 35/13 \\
Sersic 159-03   &2.0    &0.035          &0.026    & 3.8/11 \\
3C 295          &1.7    &0.22           &0.030    & 4.1/11 \\
\enddata
\end{deluxetable}

\clearpage
\begin{deluxetable}{llcll}
\rotate
\tablecaption{Observed and Predicted Scaling Relations for Galaxy Groups
and Clusters
\label{tbl2}}
\tablewidth{0pt}
\tablehead{
\colhead{Physical Quantity} &
\colhead{Observation}&
\colhead{Reference}   &
\colhead{Conduction Model} &
\colhead{Mixing Model}
}
\startdata
Size scale $r_s$         &$T^{0.5}$
                         & 1,2,3
                         &$\qquad$ ...
                         &$\qquad$ ... \\

Total mass  $M$          &$T^{1.7\sim1.9}$
                         & 3,4
                         &$\qquad$ ...
                         &$\qquad$ ... \\

Electron density  $n_e$  &...
                         &...
                         &$T$ ($T\simgt2$ keV)
                         &$T^{0.5}$ ($T\simgt2$ keV) \\

                         &...
                         &...
                         &$T^{1.6\sim1.8}$ ($T\simlt1$ keV)
                         &$T^{1.7\sim2.0}$ ($T\simlt1$ keV) \\

X-ray luminosity $L_X$   &$T^{2.5\sim3}$ ($T\simgt2$ keV)
                         & 5,6,7,8,9
                         &$T^{4}$ ($T\simgt2$ keV)
                         &$T^{3}$ ($T\simgt2$ keV) \\

                         &$T^{4\sim5}$ ($T\simlt1$ keV)
                         & 10
                         &$T^{4}$ ($T\simlt1$ keV)
                         &$T^{4.2\sim4.5}$ ($T\simlt1$ keV) \\

Entropy $S$
                         &$T^{0.6\sim0.7}$ ($T\simgt2$ keV)
                         & 11,12,13,14
                         &$T^{0.3}$ ($T\simgt2$ keV)
                         &$T^{0.7}$ ($T\simgt2$ keV) \\

                         &$T^{-0.7\sim0.2}$ ($T\simlt1$ keV)
                         & 14,15
                         &$T^{-0.2 \sim 0  }$ ($T\simlt1$ keV)
                         &$T^{-0.3 \sim -0.1}$ ($T\simlt1$ keV) \\

Gas mass fraction $f_g$
                         &$T^{0\sim0.3}$ ($T\simgt2$ keV)
                         & 2,3,16
                         &$T^{0.7}$ ($T\simgt2$ keV)
                         &$T^{0.2}$ ($T\simgt2$ keV) \\

                         &$T^{2\sim3}$ ($T\simlt1$ keV)
                         & 2,3,16
                         &$T^{1.3\sim1.5}$ ($T\simlt1$ keV)
                         &$T^{1.4\sim1.7}$ ($T\simlt1$ keV) \\

\enddata

\tablerefs{
 (1) \citealt{vik99};  (2) \citealt{moh99};  (3) \citealt{san03};
 (4) \citealt{fin01};  (5) \citealt{whi97};  (6) \citealt{mar98};
 (7) \citealt{all98};  (8) \citealt{wu99} ; (10) \citealt{hel00};
(11) \citealt{pon99}; (12) \citealt{llo00}; (13) \citealt{pra03};
(14) \citealt{pon03}; (15) \citealt{fin02}; (16) \citealt{rei01}
}
\end{deluxetable}

\clearpage

\end{document}